\documentclass{PoS}
\newcommand{\kms}{km~s$^{-1}$}
\title{The expansion of SN\,2008iz in M82}

\ShortTitle{SN\,2008iz in M82}

\author{\speaker{A.~Brunthaler}$^{a}$,  I.~Mart\'i-Vidal$^{a}$, K. M.~Menten$^a$, M. J.~Reid$^b$, C.~Henkel$^{a}$, G. C.~Bower$^{c}$, H.~Falcke$^{d,e}$, R. J.~Beswick$^{f}$, T. W. B.~Muxlow$^{f}$, and D. M.~Fenech$^{g}$\\
       \llap{$^a$} Max-Planck-Institut f\"ur Radioastronomie (MPIfR), Auf dem H\"ugel 69, 53121 Bonn, Germany\\     
       \llap{$^b$} Harvard Smithsonian Center for Astrophysics, 60 Garden Street, Cambridge, MA 02138, USA\\
\llap{$^c$} UC Berkeley, 601 Campbell Hall, Astronomy Department \& Radio
              Astronomy Lab, Berkeley, CA 94720, USA \\ 
\llap{$^d$} Department of Astrophysics, Radboud Universiteit Nijmegen, Postbus 9010, 6500 GL Nijmegen, the Netherlands\\
\llap{$^e$} ASTRON, Postbus 2, 7990 AA Dwingeloo, the Netherlands\\
\llap{$^f$} Jodrell Bank Centre for Astrophysics, School of Physics and Astronomy, The University of Manchester, Oxford Road, Manchester M13 9PL\\
\llap{$^g$} Department of Physics and Astronomy, University College London, Gower Street, London WC1E 6BT\\
E-mail:\email{brunthal@mpifr-bonn.mpg.de}, \email{imartiv@mpifr-bonn.mpg.de}, \email{kmenten@mpifr-bonn.mpg.de}, \email{reid@cfa.harvard.edu}, \email{p220hen@mpifr-bonn.mpg.de}, \email{gbower@astro.berkeley.edu}, \email{H.Falcke@astro.ru.nl}, \email{rbeswick@jb.man.ac.uk}, \email{twbm@jb.man.ac.uk},\email{dmf@star.ucl.ac.uk}
}

\abstract{We present first results from the ongoing radio monitoring of 
SN\,2008iz in M82. The VLBI images reveal a shell-like structure with circular 
symmetry, which expands in a self-similar way. There is strong evidence of a 
compact component with a steep spectrum at the center of the shell. The 
expansion curve obtained from our VLBI observations is marginally decelerated 
($m = 0.89$) and can be modeled simultaneously with the available 
radio light curves. While the results of this simultaneous fitting are not 
conclusive (i.e. different combinations of values of the 
magnetic field, CSM density profile, and electron energy distribution, provide 
fits to the available data with similar quality), additional observations should 
allow a more robust and detailed modeling.}

\FullConference{10th European VLBI Network Symposium and EVN Users Meeting: VLBI and the new generation of radio arrays\\
		September 20-24, 2010\\
		Manchester Uk}

\begin{document}

\section{Introduction}
Radio supernovae are rare events. So far only about two dozen have been detected,
the majority of which were relatively distant or quite weak, making them
difficult to study in great detail (see \cite{Bartel2009} for a recent 
review). To date, the best known example is SN\,1993J in M81, which 
has been studied extensively \cite{MartiPaperII, BietenEVN} due to the close distance 
of only 3.6 Mpc. The 
recent discovery of SN\,2008iz during High Sensitivity Array observations of 
water masers in M82 \cite{BrunthalerLC, BrunthalerCBET} offers the
possibility to study another supernova at a very similar distance in great
detail and to make a comparison to SN\,1993J. SN\,2008iz has been only detected 
in the radio range, probably because it exploded in or behind a very 
dense molecular cloud. Indeed, $^{12}$CO (J=2$\rightarrow$1) line observations 
show a dense cloud toward the position of SN\,2008iz with a H$_2$ column 
density of $\sim 5.4\times 10^{22}$ cm$^{-2}$ \cite{Brunthaler2010}. This 
explains the lack of optical, infrared \cite{Fraser2009}, and X-ray detections
\cite{Brunthaler2010}, despite sensitive
searches. 


\section{Observations}

After the discovery of SN\,2008iz, we initiated several projects to follow 
the evolution with the VLA (AB1328) from 1.4--43 GHz, the VLBA (BB272) from 
1.6--15 GHz, the eEVN at 1.7 GHz (RB003) and later with the VLBA + Effelsberg 
(BB277) at 1.6--8.4 GHz. Fortunately, M82 was frequently observed between 2007 
and 2009 as a flux calibrator source in a monitoring campaign of intraday 
variable sources with the Urumqui telescope at 5 GHz. These single dish 
observations were used to extract a well sampled 5 GHz lightcurve of SN\,2008iz 
\cite{Marchili2010}. The detection of SN\,2008iz lead also to intense 
monitoring of M82 with MERLIN, and resulted in the discovery of another 
new radio source \cite{Muxlow2009, Muxlow2010, Beswick2010}. 

Our first VLBI images at 22 GHz show a small ring, expanding at 
$\sim$21000~\kms\,\cite{Brunthaler2010}, making it one of the fastest radio 
supernovae discovered so far. The VLA radio spectrum of SN\,2008iz from the  
observation on 2009 April 27 shows a broken power-law with a spectral index 
of -1.08$\pm$0.08 in the optically thin part, and a turnover frequency of 
1.51$\pm$0.09 GHz \cite{Brunthaler2010}.  In our later epochs the VLA moved 
to C and D configuration making it more difficult to separate SN\,2008iz
from the strong extended emission of M82 at frequencies below 22
GHz. To extract the lightcurve at frequencies below 22 GHz, one has to make a
pre-explosion model of M82 with the same resolution (which is possible due to
the wealth of observations of M82 at all frequencies in the VLA archive) and 
subtract this model from the emission seen in our observations (current
work in progress). 

The single dish lightcurve has allowed us to obtain information on the 
precursor mass-loss rate, the strength of the magnetic field in the radiating 
region, the explosion date, and the deceleration of the expanding shock 
\cite{Marchili2010}. The expansion velocity from the VLBI observations,
combined with an estimate of the deceleration from the 5 GHz lightcurve yields
an explosion date in mid February 2008. 



\section{Modeling the SN\,2008iz radio data}

The standard Chevalier model of radio emission from supernovae 
\cite{Chevalier1982a, Chevalier1982b} describes the supernova ejecta 
interacting with the circumstellar medium (CSM) as a spherically-symmetric
and self-similar expanding shock, consisting of a contact discontinuity, 
a reverse shock, and a forward shock, which extends into the CSM. The 
synchrotron radio emission is assumed to be produced in the shocked CSM 
region and, therefore, the structure of a radio supernova (RSN) should be 
shell-like. Indeed, shell-like structures, and eventually their self-similar 
expansion, have been reported for several RSNe: e.g., SN\,1993J (e.g., 
\cite{MarcaideNature}), SN\,1986J (e.g., \cite{PerezTorres86J}), SN\,1979C 
\cite{Bartel79C} and SN\,2008iz \cite{Brunthaler2010}, although strong 
inhomogeneities and deformations in the shells of some sources (e.g., 
SN\,1979C and SN\,1986J) have been reported, possibly due to large 
anisotropies in their CSM.

\subsection{Expansion curve and discovery of a central component}

\begin{figure}
\centering
\includegraphics[width=10cm]{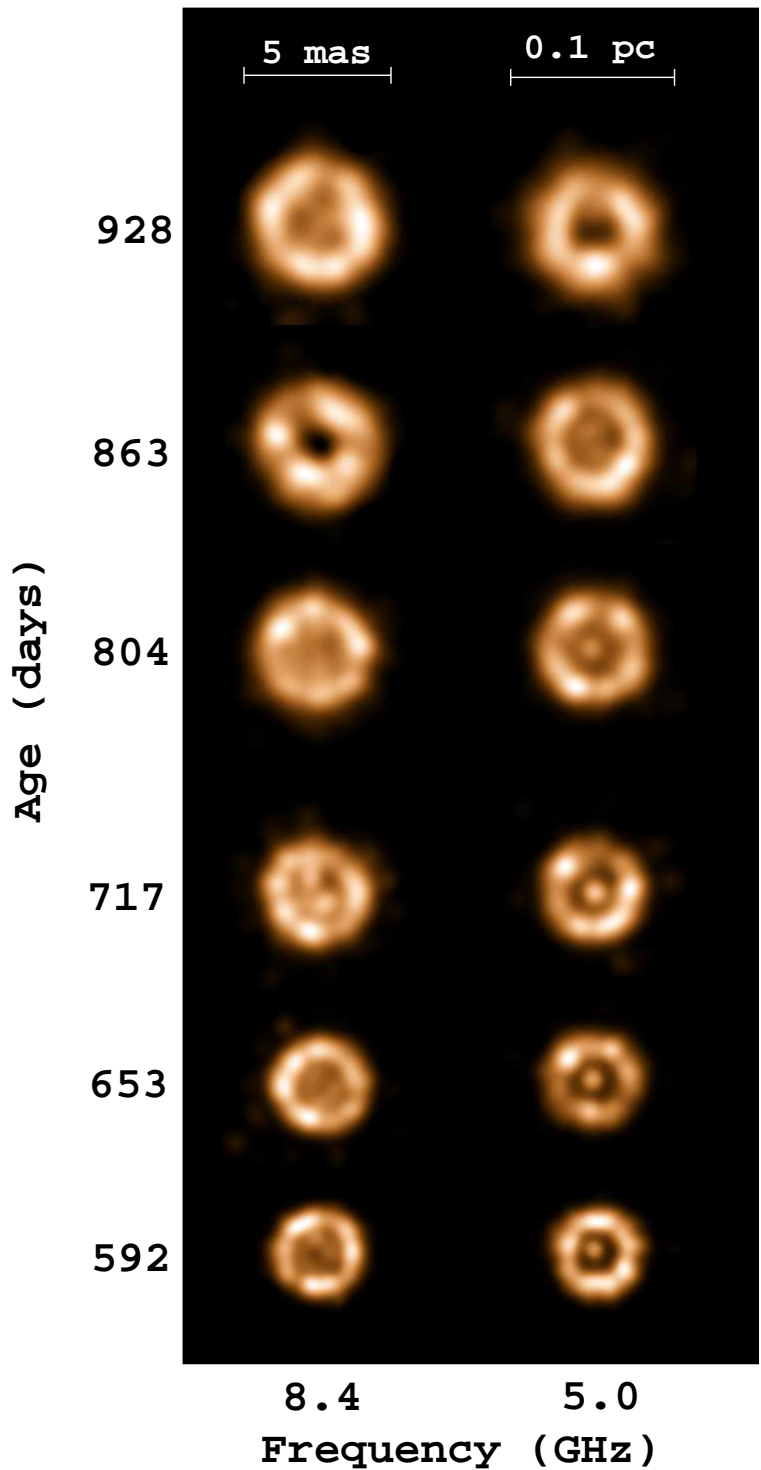}
\caption{VLBI images of SN\,2008iz at 8.4, and 5 GHz.
Emission intensity is shown as linear color scale, running from 0 Jy/beam 
(black) to the peak intensity of each image (white). CLEAN components are 
convolved with a {\em dynamic beam}, i.e., a Gaussian of FWHM equal to 1/3 of 
the shell size at each epoch. {\em Age} is time since 20 February 2008. Note that the
Effelsberg telescope was missing in the last two epochs.}
\label{Composite}
\end{figure}

The radio structure of SN\,2008iz is remarkably circularly symmetric, and its
self-similar expansion has been detected \cite{Brunthaler2010} and monitored 
with VLBI at several frequencies (these proceedings). In Fig. \ref{Composite} 
we show a composite of the images obtained from our observations at 8.4 and 5 
GHz, analyzed to date. A {\em dynamic beam} (see, e.g. \cite{MarcaideScience}),
equal to 1/3 times the shell radius, has been used to convolve the CLEAN model 
components in all epochs. The shell-like structure of SN\,2008iz can be readily
seen in all cases. Remarkably, there is strong evidence of a compact source at 
the center of the shell, which is detected in some epochs at 5 GHz, but not at 
higher frequencies.  Such a compact component was detected also  in the center 
of the SN\,1986J radio shell \cite{BietenPulsar}, which the authors identified
as either pulsar emission or related to accretion onto a black hole. Note that
the Effelsberg telescope, which provides the longest and most sensitive baselines, 
had technical difficulties in the last two epochs. Therefore these two epochs have
a significantly lower sensitivity and angular resolution, which might explain why 
the central component is not visible anymore. Our continued monitoring of SN\,2008iz
will certainly solve this issue and a detailed discussion of the steep-spectrum 
compact component discovered in the center of our SN\,2008iz VLBI images will be 
published in a forthcoming paper.

\subsection{Simultaneous fit of expansion and radio light curves}

The Chevalier model \cite{Chevalier1982a, Chevalier1982b} can be used to 
relate the parameters of the model radio 
light curves to those of the expansion curve, by means of simple analytic 
expressions (e.g., \cite{WeilerModel}). The 5\,GHz light curve of SN\,2008iz, 
taken with the Urumqi telescope \cite{Marchili2010} was fitted with the model 
from \cite{WeilerModel}. An expansion index\footnote{The expansion index, 
$m$ is such that $R \propto t^m$, $R$ being  the shell radius and $t$ the time 
after explosion} of $m = 0.89$ was then derived (or {\em predicted}) by 
\cite{Marchili2010} to properly explain their data. It is worth noticing that 
the expansion index reported in \cite{Marchili2010} fits remarkably well to 
our expansion curve, obtained from the VLBI observations recently analyzed.

\begin{figure}
\centering
\includegraphics[width=17cm]{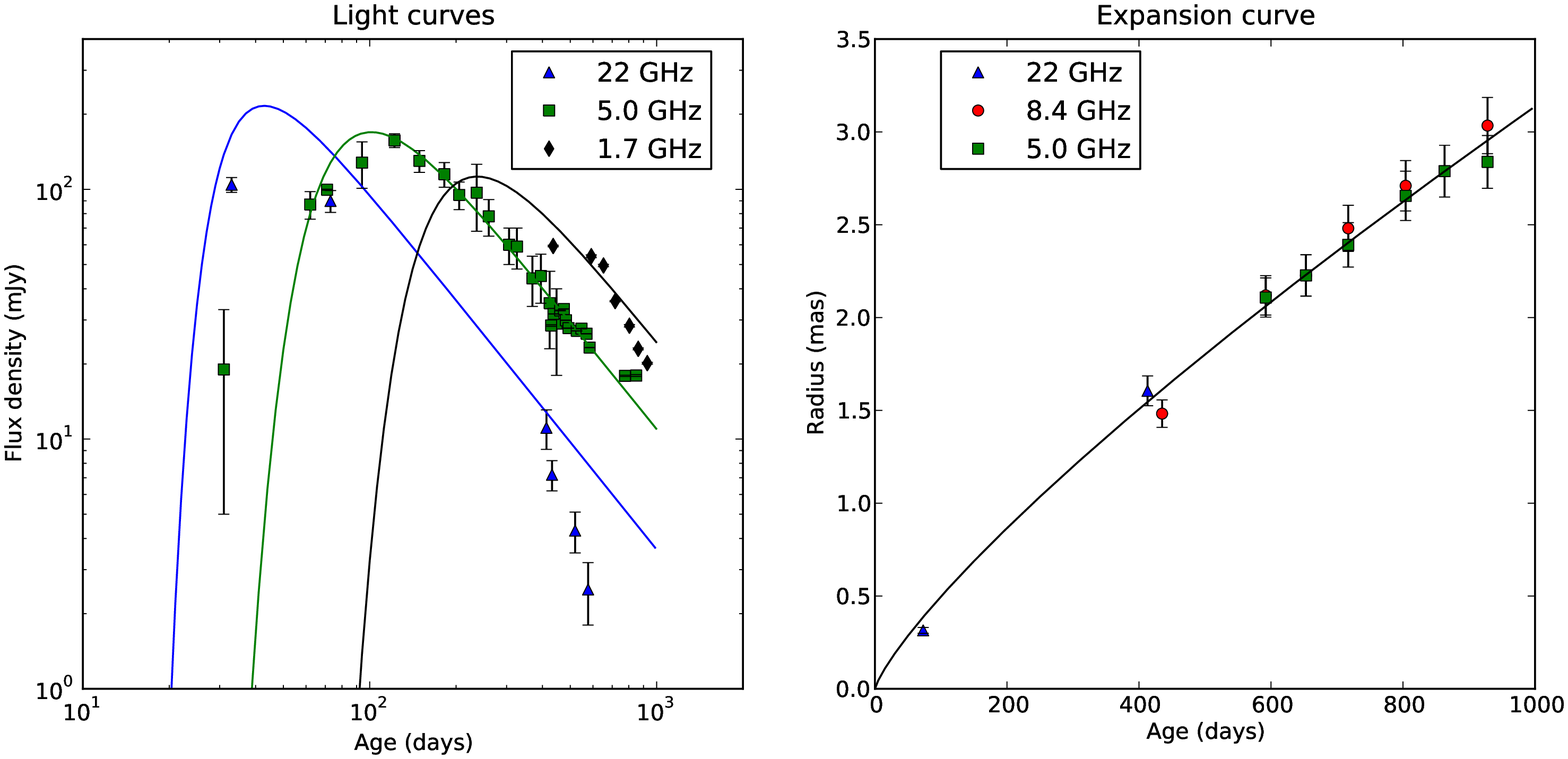}
\caption{Left, radio light curves of SN\,2008iz, taken with the Urumqi 
telescope and Merlin (data at 5 GHz; \cite{Marchili2010, Muxlow2010}), the VLA 
(22 GHz; \cite{BrunthalerLC}), and the VLBA + Effelsberg observations (1.7 
GHz). Right, expansion curve obtained from our VLBI data. Lines correspond to 
our (preliminary) simultaneous fit to all data. {\em Age} is time since 20 
February 2008.}
\label{SimulFit}
\end{figure}

Although the model of \cite{WeilerModel} has been successfully used in 
the modeling of several RSN light curves, it neglects the radiative losses 
of the electrons, which can notably affect the evolving flux density. Therefore,
incorrect estimates of the model parameters can be obtained when using only 
this model. The case of SN\,1993J is an excellent example of this: if radiative 
losses are not considered, there is strong evidence of a CSM density profile 
shallower than that corresponding to a standard stellar wind\footnote{The 
density profile of the CSM is modeled as $\rho \propto r^{-s}$, $r$ denoting
the distance to the explosion center and $s=2$ (for a standard stellar wind).} 
(e.g., \cite{Miodus93J, Weiler2007}); however, if radiative cooling is 
properly introduced in the model, the data turn out to be in excellent
agreement with a standard wind \cite{Fransson93J, MartiPaperII}. 

We have simultaneously modeled the VLBI expansion curve, and the available 
flux-density observations of SN\,2008iz, using the model described in 
\cite{MartiPaperII}, which takes into account electron cooling and was used 
to successfully model all the available radio data of SN\,1993J. Since the 
radio light curves of SN\,2008iz are not nearly as complete as those of 
SN\,1993J, we had to make several assumptions for the model, fixing some 
parameters that were left free in the fit to the SN\,1993J radio data.

We show in Fig. \ref{SimulFit} our preliminary simultaneous fit to the radio 
light curves and the expansion curve for SN\,2008iz. We use an expansion index 
of $m=0.89$, which properly describes both, the expansion curve and the 
flux-density decay rate. However, different combinations of values of the 
magnetic field, CSM density profile, and electron energy distribution, provide 
fits with similar quality, and a more detailed analysis (together with the 
inclusion of additional data points) is necessary to arrive at more robust 
results. For instance, a magnetic field as low as $\sim2$ G at day 5 after 
explosion, together with an electron energy index\footnote{The number of 
electrons is $N \propto E^{-p}$} of $p = 3$ and a CSM resulting from a 
standard stellar wind (i.e., $s = 2$) fit the data acceptably (see Fig. 
\ref{SimulFit}). However, a very large magnetic field ($\sim 100$ G at day 
5) with a smaller energy index ($p = 2.6$) and a CSM profile steeper than 
that of a standard stellar wind ($s = 2.4$) lead to a fit of similar quality. 
This last possibility, however, implies a magnetic field much larger than 
that derived from particle-field energy equipartition \cite{Marchili2010}, 
but should not be discarded using only the equipartition argument.

We notice that the latest flux-density measurements at 22 GHz decrease
much faster in time than predicted by the model. However, the same model
fits the flux-density evolution at 5 GHz acceptably. This enhanced
flux-density decay at 22 GHz (but not at 5 GHz) could be indicative of a
high-energy cutoff in the relativistic electron population. A detailed
analysis of the light curves shown in Fig. \ref{SimulFit} will be
reported elsewhere.

\section{Summary}

We have observed SN\,2008iz with VLBI at several frequencies and epochs. 
Monitoring of this supernova with VLBI, and the EVLA, is still in progress. 
The VLBI images reveal a shell-like structure with circular symmetry, which 
expands in a self-similar way. There is strong evidence of a compact component 
with a steep spectrum at the center of the shell. A discussion on this 
component will be given in a forthcoming publication. The expansion curve 
obtained from our VLBI observations is marginally decelerated ($m = 0.89$) and 
can be modeled simultaneously with the available radio light curves. The 
results of this simultaneous fit are not conclusive, due to the lack of data, 
but addition of new size measurements, and flux densities, will allow a more 
detailed modeling.

\end{document}